\documentclass[useAMS,usenatbib,a4]{mn2e}
\usepackage{epstopdf}
\usepackage{ulem}
\usepackage{amssymb}
\usepackage{natbib}
\usepackage{times}
\usepackage{graphicx}
\usepackage[usenames,dvipsnames]{pstricks}
\usepackage[colorlinks=true,linkcolor=blue,citecolor=blue]{hyperref}
 \usepackage{psfrag}
 
\def\aj{AJ}%% Astronomical Journal
\def\araa{ARA\&A}%% Annual Review of Astron and Astrophys
\def\apj{ApJ}%% Astrophysical Journal
\def\apjl{ApJ}%% Astrophysical Journal, Letters
\def\apjs{ApJS}%% Astrophysical Journal, Supplement
\def\aap{A\&A}%% Astronomy and Astrophysics
\def\aapr{A\&A~Rev.}%% Astronomy and Astrophysics Reviews
\def\jcap{J. Cosmology Astropart. Phys.}%% Journal of Cosmology and Astroparticle Physics
\def\mnras{MNRAS}%% Monthly Notices of the RAS
\def\prd{Phys.~Rev.~D}%% Physical Review D
\def\pasj{PASJ}%% Publications of the ASJ
\def\physscr{Phys.~Scr}%% Physica Scripta
\newcommand{\be}{\begin{equation}}
\newcommand{\ee}{\end{equation}}
\newcommand{\bary}{\begin{eqnarray}}
\newcommand{\eary}{\end{eqnarray}}
\newcommand{\en}{E_\nu}

\def\bi{\begin{itemize}}
\def\ei{\end{itemize}}
\def\lsim{\mathrel{\rlap{\lower3pt\hbox{\hskip1pt$\sim$}}
     \raise1pt\hbox{$<$}}} %less than or approx. symbol
\def\gsim{\mathrel{\rlap{\lower3pt\hbox{\hskip1pt$\sim$}}
     \raise1pt\hbox{$>$}}} %greater than or approx. symbol

\begin{document}
\title[Study of the PeV Neutrino, $\gamma$-rays and UHECRs  around The Lobes of Centaurus A]
{Study of the PeV Neutrino, $\gamma$-rays and UHECRs \\  around The Lobes of Centaurus A}
% Force line breaks with \\
%\date{\today} % It is always \today, today,
             %  but any date may be explicitly specified
\author[N. Fraija et al.]
{N. Fraija$^{1}$\thanks{E-mail:nifraija@astro.unam.mx},   E.~Aguilar-Ruiz$^1$,  A.~Galv\'an-G\'amez$^1$,  A.~Marinelli$^2$ and J. A. de Diego$^1$\\
 $^1$Instituto de Astronom\' ia, Universidad Nacional Aut\'onoma de M\'exico, Circuito Exterior, C.U., A. Postal 70-264, 04510 M\'exico City, M\'exico\\
 $^2$ I.N.F.N. \& Physics Institute Polo Fibonacci Largo B. Pontecorvo, 3 - 56127 Pisa, Italy }%\\
\maketitle
\date{\today} 	
\begin{abstract}
Pierre Auger observatory reported the distribution of arrival directions of the highest energy cosmic rays.  These events were collected in 10 years of operations with declinations  between -90$^\circ$ and +45$^\circ$.   The IceCube neutrino telescope reported the detection of 82 extraterrestrial neutrinos in the High-Energy Starting Events catalog.  The highest-energy neutrino event (IC35)  reported in this catalog  had an energy of 2004$^{+236}_{-262}$ TeV and  its reconstruction was centered at RA=$208.4^\circ$ and DEC=$-55.8^\circ$ (J2000).    Being Centaurus A  the nearest radio-loud active Galactic nucleus and one of the potential candidates for accelerating cosmic rays up to $\sim\, 10^{20}$ eV,   we show that the ultra-high-energy cosmic rays (UHECRs)  around the direction of Centaurus A (15$^\circ$ radius) could be accelerated inside the giant lobes.    Studying the composition of UHECRs through the photo-disintegration processes,   the maximum energy that these can reach in the giant lobes and the average deflecting angles that nuclei undergo due to Galactic and extragalactic magnetic fields, it is shown that the most promise candidates of UHECR composition are light nuclei such as carbon/nitrogen nuclei.  Considering the interactions of the relativistic carbon nuclei with photon fields (extragalactic background light and synchrotron) and materials  (inside and outside the lobes), we do not find enough evidence to associate the IC35 event with the UHECRs reported by Pierre Auger observatory.
%
%we found that  the IC35 event can not be spatially associated with these UHECRs and therefore could not be generated either inside or around the giant lobes. 
%
\end{abstract}
\begin{keywords}
Galaxies: active -- Galaxies: individual (Centaurus A)  --  Physical data and processes: acceleration of particles  --- Physical data and processes: radiation mechanism: nonthermal -- Neutrinos
\end{keywords}
\section{Introduction}
At a distance of $d_z$=3.8 Mpc, Centaurus A   \cite[Cen A, classified as Fanaroff $\&$ Riley Class I;][]{1974MNRAS.167P..31F} is one of the closest and hence one of the best studied extragalactic sources. This source has  misaligned bipolar jets \citep[see, e.g.] [and reference therein]{2006PASJ...58..211H} and two giant lobes (oriented primarily in the north-south direction).   Cen A has been extensively observed  from radio wavelengths to X-rays \citep{1975ApJ...199L.139W,1976ApJ...206L..45M,1970ApJ...161L...1B,1981ApJ...244..429B}  and   MeV to  TeV $\gamma$-rays \citep{2003ApJ...593..169H,1999APh....11..221S,2009ApJ...695L..40A, 2010ApJ...719.1433A,2011A&A...531A..30R}.   A large number of scenarios has been discussed to interpret the GeV-TeV energy range such as one-zone synchrotron self-Compton model  \citep[SSC;][]{2010ApJ...719.1433A},   external inverse Compton \citep{2010Sci...328..725A},  photo-disintegration of heavy nuclei \citep{2013arXiv1312.7440K},  proton-photon (p$\gamma$) \citep{2012ApJ...753...40F, 2012PhRvD..85d3012S, 2014A&A...562A..12P} and proton-proton (pp) interactions \citep{2009NJPh...11f5016D, 2014ApJ...783...44F}.  \\ 
The Pierre Auger Observatory (PAO) located in Malargue, Argentina, reported early the distribution of arrival directions of 27 UHECRs, from 2004 January 1$^{\rm th}$ up to 2007 August 31,  with energies larger than 57 EeV \citep{2007Sci...318..938P,  2008APh....29..188P}.  A statistically significant correlation was claimed  between these UHECR events and the distributed galaxies reported in the $12^{\rm th}$ Veron-Cetty and Veron \citep[VCV;][]{2006A&A...455..773V} catalog of AGN.    Recently, \cite{2015ApJ...804...15A} reported the distribution of arrival directions of the highest energy cosmic rays over 10 years of operations, from 2004 January 1$^{\rm th}$ up to 2014 March 31, associating  some UHECRs with energies larger than 58 EeV around the direction centered in Cen A.  Based on the analysis performed by the Pierre Auger collaboration about the number of UHECRs bigger around Cen A than the rest of the sky \citep{2007Sci...318..938P, 2008APh....29..188P,  2015ApJ...804...15A},  this radio galaxy has been widely proposed as a potential source for emitting the UHECR events \cite[e.g.][]{2008JETPL..87..461G,2009ApJ...693.1261M, 2009NJPh...11f5016D, 2012ApJ...753...40F}.   Requiring a satisfactory description of the $\gamma$-ray spectra with hadronic models, some authors have extrapolated the power law of accelerated protons up to energies of $\sim10^{20}$ eV in order to explain the number of UHECRs observed by PAO  \citep{2014ApJ...783...44F, 2014A&A...562A..12P, 2012PhRvD..85d3012S}. \\
The IceCube experiment located at the South Pole \citep{2006APh....26..155I, 2004APh....20..507A} reported evidence for extragalactic neutrinos in the TeV - PeV energy range.  The collaboration reported, first,  two PeV neutrino events \citep{2013PhRvL.111b1103A}, next  they added 26 events \citep{2013Sci...342E...1I},  later 9 more events \citep{2014PhRvL.113j1101A}, and finally, 17 events were included to this long list. In total, 54 neutrino events were reported in the High-Energy Starting Events (HESE)\footnote{http://icecube.wisc.edu/science/data/HE-nu-2010-2014} catalog \citep{2015arXiv151005223T}.  Recently, 28 neutrinos events (from 2014 to 2016) were added to this catalog \citep{2017arXiv171001191I}. The IC35 neutrino event centered at RA=$208.4^\circ$ and DEC=$-55.8^\circ$ (J2000) with  a median angular error of 15.9$^\circ$ had the highest energy reported in this catalog which corresponds to 2004$^{+236}_{-262}$ TeV.   Using six years (from 2009 to 2015) of data  and charged current muon neutrino events, IceCube reported a complementary measurement of 29 neutrino events in the Northern Hemisphere  \citep{2016ApJ...833....3A}.  Aiming for the identification of an electromagnetic counterpart with possible multi-messengers,  the IceCube neutrino telescope  has established a system of real-time notifications that rapidly alert the astronomical community of the position of an astrophysical neutrino candidate \citep{2017APh....92...30A}. From the beginning of the program, in April 2016 through November 2017,  12 notifications have been issued for well-reconstructed muon tracks produced in this telescope.\\
\\
The IC35 neutrino event has been studied in various contexts \citep[i.e. see; ][]{2016MNRAS.457.3582P, 2014MNRAS.443..474P, 2016PhRvL.116o1101W, 2016NatPh..12..807K}.  For instance,  \cite{2016NatPh..12..807K} revised the second catalog of AGN reported by Fermi Large Area Telescope \citep[LAT, 2LAC,][]{2011ApJ...743..171A} and found 20 astrophysical sources associated to the median angular error of this event.  The nearest AGN in the field of view of this event is Cen A.  Using the Fermi-LAT data collected from  the core of Cen A, \cite{2013APh....48...30S} normalized the neutrino flux and found  that the density required to observe  high-energy neutrinos would have to be or the order of 10 - 100 cm$^{-3}$, thus discarding Cen A as a possible candidate to emit the IC35 neutrino event.  The dominant blazar associated to the median angular error of IC35 neutrino event is PKS B1424-418 which was first discarded by \citet{2014MNRAS.443..474P} due to its low $\gamma$-ray emission. However,  this source presented an outstanding flare  in temporal coincidence with the IC35 event, beginning in summer 2012 and lasting for almost one year.   \cite{2016NatPh..12..807K} considered a photo-hadronic emission model and linked this event with the giant outburst presented by PKS B1424-418.\\  
\\
In this paper, we study the possible association of the PeV-neutrino event (IC35) reported in the HESE catalog  with the UHECR events collected by PAO between 2004 January 1th and 2014 March 31. This study was presented around the extended lobes of Cen A and their closest galaxies ($\lesssim$ 4 Mpc).   For this analysis, we will focus on different  places of acceleration and their respective timescales (inside and outside the extended lobes), the chemical composition of UHECRs and the interactions among the relativistic carbon nuclei with photon fields (cosmic microwave background, extragalactic background light (infrared, optical and ultraviolet) and synchrotron) and distinct materials  (inside and outside the extended lobes).  In addition, we present a description of the faint $\gamma$-ray fluxes reported by Fermi Collaboration through the hadronic interactions among the relativistic carbon nuclei with thermal density material inside the giant lobes.  The paper is arranged as follows.  In Section 2 we introduce relevant information and previous works about UHECRs,  neutrinos and $\gamma$-rays in the giant lobes of Cen A.  In Section 3 we present our analysis done about UHECRs,  neutrinos and $\gamma$-rays inside the giant lobes of Cen A. In Section 4 we present an analysis requiring  the closest Galaxies around Cen A  and the hadronic interactions in their paths to Earth,  and in Section 5 we show the conclusions.  We hereafter present the equations in natural units and also the redshift z$=0.00183(2)\simeq 0$.
%
%we investigate the possible association of UHECRs with the giant lobes of Cen A,  the UHECR composition and also the conditions so that the PeV-neutrino event (IC35) reported in the HESE catalog could be associated to UHECRs detected by PAO around the giant radio lobes of Cen A.
%
\section{UHE Cosmic Ray,  Neutrino and $\gamma$-rays in the extended Lobes of Cen A}
Cen A has two giant lobes which are subtended $\sim 10^\circ$ ($\sim$ 600 kpc in projection) on the sky \citep{1998A&ARv...8..237I}.  These extended lobes are connected to the energy supply jet which injects  nuclei/protons to them  \citep{2009MNRAS.393.1041H}.    The  non-thermal and the upper limit thermal pressure can be estimated by $p_{\rm nth}\simeq (U_e+U_B+U_{A})$  and p$_{\rm th}=n_p\,T$, respectively, where $U_e$ is the electron energy density, $U_B$ is the magnetic energy density,  $U_A\sim 2\, L_{A}\, t_{\rm lobe}/V$ is the accelerated-nuclei/proton energy density with luminosity $L_A$, $n_p$ is the particle density,  $T$ is the temperature and $V$ is the volume of the giant lobes.  Using the relationship between the magnetic field and the electron energy densities, $U_e=\lambda_{\rm e,B}\,U_B$ \citep{2010Sci...328..725A, 2014ApJ...783...44F},  the non-thermal pressure and the total energy can be written as
\be\label{pressure}
p_{\rm nth}\simeq U_B(1+\lambda_{\rm e,B})+ 2\, L_A\frac{t_{\rm lobe}}{V},
\ee
and
\be\label{Etotal}
E_{\rm tot}\simeq \,U_B\,V(1+\lambda_{\rm e,B})+ 2\, L_A\,t_{\rm lobe},
\ee
respectively.\\

\subsection{UHE Cosmic Rays}
The Pierre Auger Collaboration reported initially an excess of UHECR events very close to the  direction of Cen A \citep{2010APh....34..314A}.  They found 13 UHECRs with energies above $\geq$ 55 EeV in a circular window radio of $18^\circ$ centered on this radio galaxy. Later,  \cite{2015ApJ...804...15A} performed cross-correlations of events with the AGN in the Swift catalog.  Authors found 14 UHECR events with energies above $\geq$ 58 EeV in a circular window radio of $15^\circ$.  Considering the UHECR events reported in the direction of Cen A, we study the deflection and the composition of UHECRs as well as mechanisms and places of acceleration. Additionally, the UHECR luminosities are estimated.
\subsubsection{Deflection and composition}
UHECRs traveling from the giant lobes of Cen A to Earth are randomly deviated between the original and the observed arrival direction due to the strengths and geometries of magnetic fields: extragalactic and Galactic \citep{2008ApJ...682...29D, 2010ApJ...710.1422R}, Galactic winds \citep{2016MNRAS.458..332H, 2016MNRAS.456.1723L, 2016yCat..51500081W}, magnetic winds \citep{1958ApJ...128..664P,2011ApJ...739...60E} and magnetic turbulence \citep{2013MNRAS.436.1245F, 2008Sci...320..909R}.   For instance,  distinct strengths of extragalactic and Galactic magnetic fields produce deflections of {\small $\theta_{\rm Z,EG}\sim Z E_{\rm A,th}^{-1} B_{\rm EG}\,L_{\rm EG}^{1/2}\,l_{\rm c,EG}^{1/2}$} and {\small $\theta_{\rm Z,G}\sim Z E_{\rm A,th}^{-1} B_{\rm G}\,L_{\rm G}^{1/2}\,l_{\rm c,G}^{1/2}$}, respectively, where  Z is the atomic number,  L$_{\rm G}$  (L$_{\rm EG}$) corresponds to a Galactic (extragalactic) distance,  $l_{\rm c,G}$ ($l_{\rm c, EG}$) is the coherence length for the Galactic ($B_{\rm G}$) and extragalactic ($B_{\rm EG}$) magnetic fields, respectively,  and $E_{\rm A,th}$ is the threshold nuclei/proton energy.\\ 
Due to Cen A is located at 19.4$^\circ$ from the Galactic plane and  43.1$^\circ$  from the Galactic center, UHECRs coming from this radio galaxy could undergo large deflections in their paths ($\sim$ dozens of kpc) by the stronger magnetic field ($\sim {\rm 1-4\mu G}$) close to the Galactic plane.  Therefore, UHECRs traveling close to the Galactic plane might have  deflections as large as (or even larger than) those produced by the weak ($\sim 1\, {\rm nG}$) extragalactic magnetic field in larger trajectories. It is worth noting that the strength of Galactic magnetic field on the plane is higher than magnetic field in the halo, and the paths from this radio galaxy have to cross twice the size of our galaxy \citep{2008PhyS...78d5901F}.\\
The composition of UHECRs is not clear and it is under debate.  Studies reported by observatories as AGASA \citep{2006AJ....131.2843S}, Yakutsk \citep{2008NuPhS.175..201K} and others \citep{2007JPhG...34..401S, 2005ApJ...622..910A} supported that at the highest energies  the UHECR composition was dominated by protons.   However, others experiments such as PAO have suggested that the UHECR composition at the highest energies was dominated by heavier or mixed composition \citep{2007AN....328..614U} and recently by light nuclei \citep{2014PhRvD..90l2006A}.  Recently, this collaboration  analyzed data from December 2004 to December 2012 in order to examine the implications of the depth distributions in the atmosphere for composition of the primary cosmic rays. They evaluated several  hadronic interaction models and analyzed the quality of fit in the whole energy range. They found that whereas data were not adjusted by a mix of proton (p) or iron (Fe) nuclei, light nuclei such as nitrogen (N) described successfully data at the highest energies.   It is worth noting that the interpretation of the results depends on the hadronic interaction model used to extrapolate the information of the cross section up to the highest energies.  Therefore, important details  about chemical composition in the extrapolated data could be misunderstood \citep{2007JCAP...12..002A}.\\
\subsubsection{Mechanisms of acceleration}
The maximum energies \citep[so-called Hillas condition;][]{1984ARA&A..22..425H}   that CRs can reach within the extended lobes are limited by their sizes (R) and the strength of magnetic fields (B).    Although the maximum  nuclei/proton energy  obtainable for relativistic shocks is far from clear, the energy adopted through the Hillas condition is too optimistic.   \cite{2014MNRAS.439.2050R} investigated the maximum energy that CRs can reach at weakly magnetized ultrarelativistic shocks.  This maximum energy was estimated from the condition that particles must be isotropized in the downstream side before the field carries them far downstream. Authors proposed that to avoid this limit, a highly disorganized field is needed on larger scales.\\
During the flaring events,  CRs could be also accelerated  up to UHEs.   The maximum particle energy of accelerated CRs depends on  the equipartition magnetic field parameter and the apparent isotropic luminosity \citep{2009NJPh...11f5016D}.\\
Other mechanisms of acceleration proposed to accelerate CRs in the giant lobes have been the magnetic reconnection and the stochastically acceleration by temperature gradient. In the magnetic reconnection framework, the free energy stored in the helical configuration can be converted to particle kinetic energy,   resulting in continuously charged particle acceleration \citep{2000ApJ...530L..77B, 2010MNRAS.408L..46G,  2013A&A...558A..19W}.   Stochastic  acceleration of CRs by  high temperatures has been proposed by \citet{2013A&A...558A..19W}.  In this mechanism,  CRs could be stochastically accelerated, being responsible for  the self-consistency between the entrainment calculations and the missing pressure in the lobes.  \cite{2009MNRAS.400..248O} investigated the acceleration of particles via the second-order Fermi process in the extended lobes. They concluded that considering this process,  the maximum energy reached by CRs was in the energy range observed by PAO.\\
\subsubsection{UHECR Luminosity}
Cen A has been proposed as a potential source for studying the UHECR events \citep{2008JETPL..87..461G,2009ApJ...693.1261M, 2009NJPh...11f5016D, 2012ApJ...753...40F}.   Recently,  \cite{2015ApJ...804...15A} reported the distribution of arrival directions of UHECRs collected with PAO in 10 years of operations. To determine the number of UHECRs,  we use the PAO  exposure,  which  for a point source is given by $\Xi\, \omega(\delta_s)/\Omega_{60}=(6.6\times10^4 \times 0.64/\pi$) km$^2$yr , where  $\omega(\delta_s)$ is an exposure correction factor for the declination of Cen A \citep{2008PhRvD..78b3007C}.   The nuclei/proton luminosity {\small$ L_{\rm A}= 4\,\pi\,d^2_z A_{\rm A}\,\int \,E_{\rm A}\,E_{\rm A}^{-\alpha_A}  dE_A\,$} above an energy $E_{\rm A,th}=58$ EeV for a simple power law  {\small $(dN/dE)_{\rm A}=A_A E^{-\alpha_A}$} can be estimated as 
\bary
L_A=  4\pi d^2_{\rm z}\frac{\epsilon_0\,\Omega} {\Xi\, \omega(\delta_s)}  \left(\frac{\alpha_{\rm A} - 1}{\alpha_{A} - 2}\right)E^{-1+\alpha_{A}}_{\rm A,th} N^{\rm obs}_{\rm cr}\,E^{2-\alpha_{\rm A}}_{\rm A},
\label{num}
\eary
where this equation is normalized with the number of UHECRs ($N^{\rm obs}_{\rm cr}$) to be found in the direction of interest. The quantity  $\epsilon_0=1\,{\rm EeV}$ corresponds to the normalization energy.\\
\subsection{Neutrinos}
CRs unavoidably interact  with external radiation fields and ambient gas producing HE neutrinos whereas they propagate through the extended lobes.
\subsubsection{Photo-hadronic interactions}
Accelerated nuclei/protons interact with the cosmic microwave background (CMB)  and extragalactic background light (EBL) photon fields by photo-hadronic interactions.   Following \citet{2014PhRvD..90b3007M}, the neutrino spectrum can be obtained by means of the nuclei/proton spectrum given by 
\be\label{pg}
E_\nu L_\nu\simeq\frac{3}{8} f_{A\gamma} E_{\rm A} L_{\rm A}\,,
\ee
where  $f_{A\gamma}\simeq {\rm min} \{\sigma^{\rm eff}_{A\gamma} n_{\gamma}\, R,\, 1\} $ \citep{ste68, PhysRevLett.78.2292} is the photomeson production efficiency, with $n_\gamma$ the photon density of CMB and EBL,  and $\sigma^{\rm eff}_{A\gamma}  \simeq \sigma_{\rm pk, A}\, \xi_{p\gamma}\,\frac{\Delta\epsilon_{pk}\,}{\epsilon_{pk}}$ with $\sigma_{\rm pk, A}\approx 5\times\,10^{-28}\,\, {\rm cm^2 A}$,   $\Delta\epsilon_{\rm pk}$=0.2 GeV, $\epsilon_{\rm pk}\simeq$ 0.3 GeV and $\xi_{p\gamma}\simeq 0.2$ per nucleon.  The characteristic neutrino energy can be roughly estimated as  
 \be
\epsilon_{\rm \nu,br}\simeq0.25\, \xi_{p\gamma}\,\delta_D^2\,(m_\Delta^2-m_p^2) \epsilon^{-1}_{\rm \gamma,br}\,,
\label{pgamma}
\ee
where $m_\Delta$ and $m_p$ are the resonance and proton masses,  $\epsilon_{\rm \gamma,br}$ is the peak photon energy of EBL and CMB, and $\delta_D$ is the Doppler factor.
\subsubsection{Hadronic interactions}
Accelerated nuclei/protons interact with the ambient internal density protons.  The neutrino spectrum can be obtained through the nuclei/proton spectrum given by 
\be\label{pp}
E_\nu L_\nu\simeq f_{Ap} E_A L_A\,,
\ee
where  $f_{Ap}\simeq {\rm min} \{\sigma_{Ap}\,n_p\, t_{\rm lobes},\,1\}$ \citep{2002MNRAS.332..215A, 2012ApJ...753...40F} is the production efficiency of this process with $\sigma_{Ap}\simeq 8\times 10^{-26} {\rm cm^2}\,\, A^{3/4}$ at $\sim$ 100 PeV, $n_p$ is the proton density and $t_{\rm lobes}$ is the age of the giant lobes. 
\vspace{1cm}
\subsubsection{Expected neutrino events}
\begin{figure}
\vspace{0.4cm}
{\centering
\resizebox*{0.5\textwidth}{0.3\textheight}
{\includegraphics{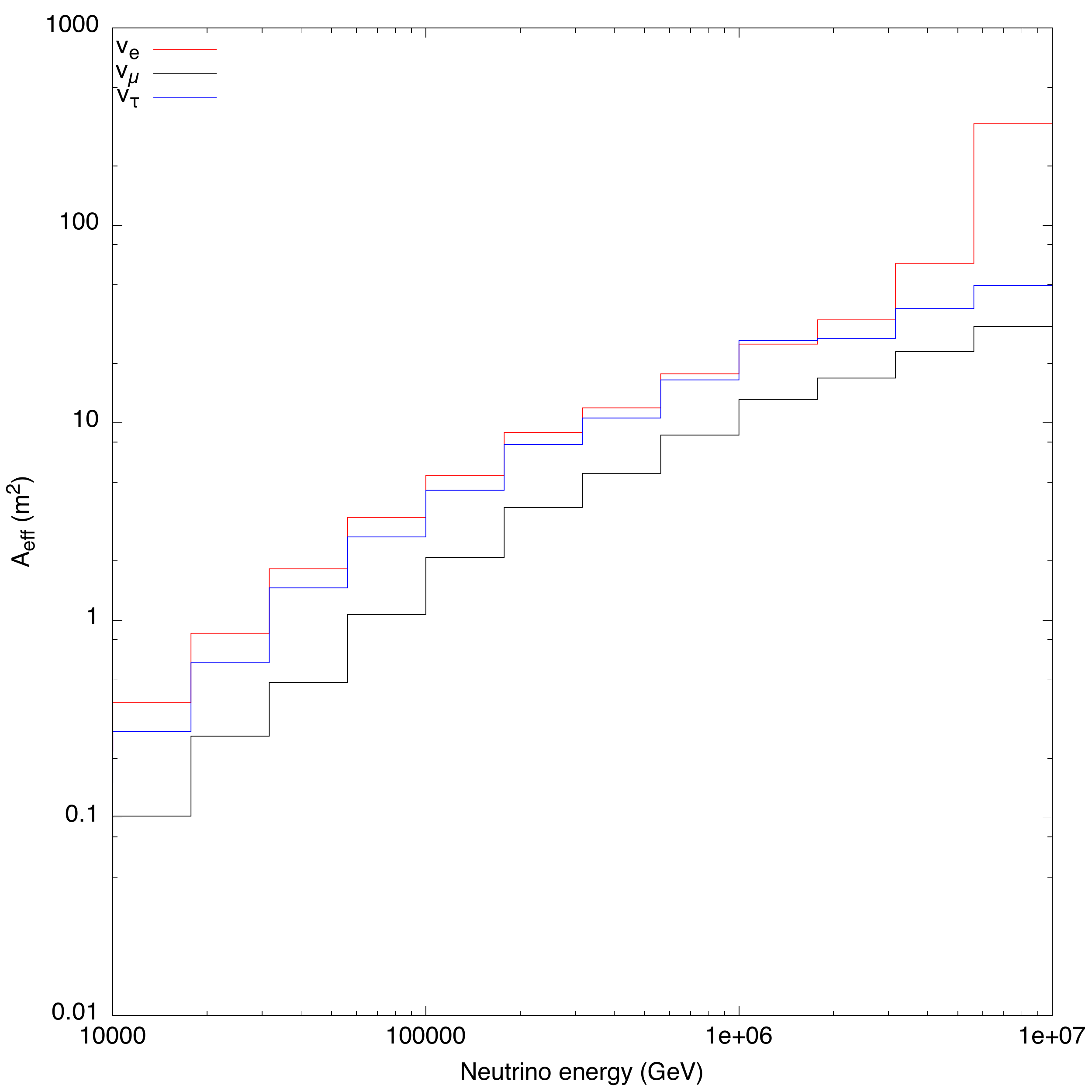}}
}
\caption{Effective area of electron, muon and tau neutrino for a point-like source in the declination of Cen A. }
\label{Aeff}
\end{figure}

The number expected of shower-like neutrino events in IceCube telescope is
\be\label{Nev}
N_{ev} \approx\,T \,\int_{E_{\rm \nu, min}}^{E_{\rm \nu, max}}\,A_{\rm eff} \, \left(\frac{dN_{\nu}}{d\en}\right)\,dE_\nu \,,
\label{numneu1}
\ee
where $T\simeq$ 6 years is the lifetime for HESE catalog, $A_{\rm eff}$ is the neutrino effective area for the shower-like events  in the declination of Cen A (see Figure \ref{Aeff}) and $dN_\nu/dE_\nu$ is the neutrino spectrum which can be written as 
\be
\frac{dN_\nu}{d\en}=A_\nu \left(\frac{E_\nu}{\rm 100\,TeV}\right)^{-\alpha_\nu}\,,
\ee
with $\alpha_\nu$ the neutrino spectral index, $E_{\rm \nu, max}$ and $E_{\rm \nu, min}$ for the maximum and minimum  neutrino energy, respectively. The normalization constant $A_\nu$ is estimated to be
\be
A_\nu\simeq\frac{L_\nu}{4\pi d^2_z \int_{E_{\rm \nu, min}}^{E_{\rm \nu, max}}\,E_\nu\, \left(\frac{E_\nu}{\rm 100\,TeV}\right)^{-\alpha_\nu}\,dE_\nu}\,.
\ee

\subsection{High-energy $\gamma$-ray photons}
The giant lobes have been detected in  radio  and $\gamma$-ray bands by the Parkes radio telescope  \citep{1993A&A...269...29J, 2000A&A...355..863A},  the Wilkinson Microwave Anisotropy Probe  \citep{2009ApJS..180..225H,2003ApJS..148...39P, 2009MNRAS.393.1041H, 2010Sci...328..725A} and Fermi-LAT   \cite{2010Sci...328..725A}.  The Fermi collaboration reported  $\gamma$-ray excesses with energies larger than $\geq$ 100 MeV for both giant lobes of Cen A \citep{2010Sci...328..725A}. The faint  $\gamma$-ray fluxes  at 5 GeV  were $F_{\rm ph}\simeq 4.6 \times 10^{-12}\,{\rm erg\,cm^{-2}\,s^{-1}}$ and $\simeq 1.2 \times 10^{-12}\,{\rm erg\,cm^{-2}\,s^{-1}}$ which correspond to photon luminosities of  $L_\gamma\simeq7.8\times10^{39}\,{\rm erg\,s^{-1}}$  and $1.9\times10^{39}\,{\rm erg\,s^{-1}}$  for the southern and northern lobes, respectively.   The Fermi-LAT collaboration  used a leptonic model to describe the electromagnetic  emission from the lobes; Radio (WMAP) data through synchrotron radiation and  Fermi-LAT data by  inverse Compton-scattered (IC) radiation from  CMB and EBL photon fields. \citet{2014ApJ...783...44F} evoked a lepto-hadronic model to describe de spectral energy distribution (SED) of the giant lobes; synchrotron radiation to explain the radio wavelength emissions  and  hadronic interactions to fit the $\gamma$-ray fluxes. \\ 
Table \ref{table1} shows the parameters derived from observations of the giant lobes.
\begin{table}
\centering
\caption{Derived parameters of the giant Lobes}\label{table1}
\begin{tabular}{ l c c c c }
 \hline
 Giant lobes & \scriptsize{} & \scriptsize{Northern} &\scriptsize{Southern} \\
 \hline 
 \hline
\scriptsize{Magnetic field ($\mu$G)}                              & \scriptsize{B}  &  \scriptsize{0.9 - 3.4} & \scriptsize{0.9 - 6.2} &   \scriptsize{(1,2)}   \\
\scriptsize{Equipartition parameter}               & \scriptsize{$\lambda_{\rm e,B}$}  & \scriptsize{4.3} & \scriptsize{1.8} &   \scriptsize{(1)}  \\
\scriptsize{Density particles  (${\rm cm^{-3}}$)}  & \scriptsize{$n_{\rm p}$}  & \scriptsize{10$^{-11}$ - 10$^{-4}$} & \scriptsize{10$^{-11}$ - 10$^{-4}$}&   \scriptsize{(3,4,5)}    \\
\scriptsize{Size of the Lobes (kpc)}                & \scriptsize{$R$}  & \scriptsize{100 - 230} & \scriptsize{100 - 220}  &   \scriptsize{(1,3)} \\ \hline
\scriptsize{Age of the lobes} & \scriptsize{}  & \scriptsize{} & \scriptsize{}  &   \scriptsize{}  \\ \hline
\scriptsize{Spectral arguments (Myr)} & \scriptsize{$t_{\rm lobe}$}  & \scriptsize{21 - 55} & \scriptsize{24 - 27}  &   \scriptsize{(2,3)}  \\
\scriptsize{Buoyancy arguments (Myr)} & \scriptsize{$t_{\rm lobe}$}  & \scriptsize{$\sim$ 600} & \scriptsize{$\sim$ 600}  &   \scriptsize{(5)}  \\
\hline
\end{tabular}
\begin{flushleft}
\scriptsize{
{References}. (1) \cite{2010Sci...328..725A} (2) \cite{2014ApJ...783...44F} (3) \cite{2009MNRAS.393.1041H}  (4) \cite{2013ApJ...766...48S}   (5) \cite{2013A&A...558A..19W}}
\end{flushleft}
\end{table}
\section{Analysis inside the extended Lobes}
\subsection{UHE Cosmic Rays}
\subsubsection{Timescales, composition and deflection}

\begin{figure}
\vspace{0.4cm}
{\centering
\resizebox*{0.5\textwidth}{0.3\textheight}
{\includegraphics{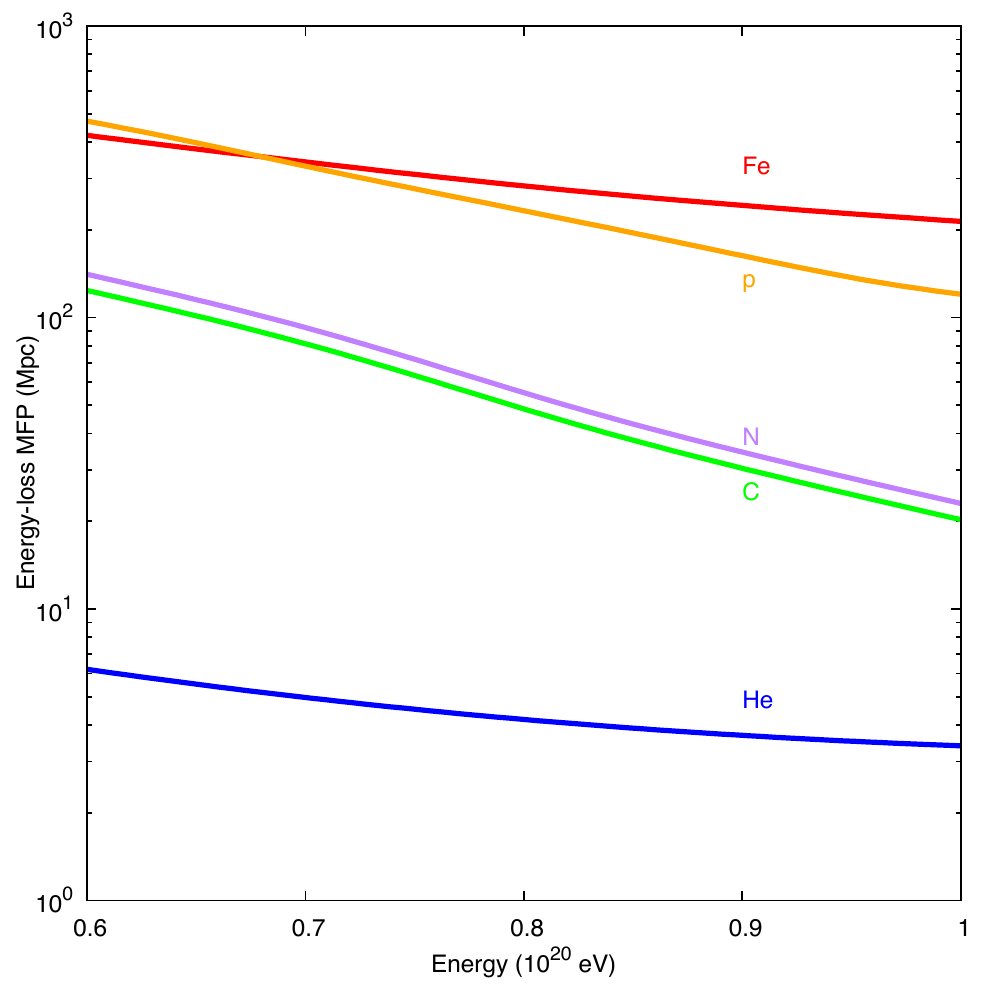}}
}
\caption{Energy-loss mean free path of UHECR proton and ions (He, C, N and Fe) due to interactions with CMB photons.}
\label{FMP}
\end{figure}

In subsection 2.1.2,  we present some mechanisms that could accelerate CRs up to UHEs.  As follows, we show that nuclei/protons can be accelerated up to ultra-relativistic energies inside the giant lobes and also are able to escape from them.  Using the values reported of the giant lobes (see Table \ref{table1}), then the  maximum energy that CR can reach by stochastic processes inside the giant lobes is 
\be\label{epmax}
E_{\rm Z, max}= 0.17\times 10^{20}\, {\rm eV}\, Z\, \left(\frac{B}{\rm 1\,\mu G}\right)\,\left(\frac{R}{\rm 100\, kpc}\right)\,\left(\frac{\beta}{0.2}\right)\,.
\ee
The value of the magnetic field $B\sim 1\, {\rm\mu G}$ in a large scale of $100\,{\rm kpc}$ agrees with the requirement needed so that CRs could be accelerated up to  UHEs \citep{2014MNRAS.439.2050R, 2010MNRAS.402..321L}.  Although, magnetic  fields on very short scales seems to be a relevant ingredient in Fermi processes at ultrarelativistic ($\beta\Gamma\gg1$) shock waves, it disfavours the acceleration efficient of particles.  Acceleration inefficiency is expected for accelerators for which $\beta\Gamma\gg1$  and magnetic  fields are immersed on short spatial scales.\\
Equation (\ref{epmax}) indicates that the maximum energies for Z$\geq$ 6 are in agreement with the highest energy cosmic rays detected by PAO around the direction of Cen A.  This equation shows that protons, helium, lithium,  beryllium and even boron can hardly be accelerated up to energies of $10^{20}$ eV or more. The acceleration timescale in the giant lobes for ultra-relativistic nuclei is
{\small
\be\label{tac}
t_{\rm Z,acc}\simeq  9.3\,{\rm Myr}\,\,\, Z^{-1}\,\eta   \,\left(\frac{E_{\rm Z, max}}{\rm 10^{20}\,{\rm eV}}\right)\,\left(\frac{B}{\rm 1\,\mu G}\right)^{-1}\, \left(\frac{\beta}{0.2}\right)^{-2}   \,,    
\ee
}
which is much smaller that the ages reported for the giant lobes (see Table \ref{table1}).   The characteristic timescale of diffusive escape for nuclei is 
{\small
\be\label{tdif}
t_{\rm Z,esc}\simeq 0.4\,{\rm Myr} \,\,\,Z\,\eta^{-1}\,\left(\frac{R}{\rm 100\, kpc}\right)^2\,\left(\frac{B}{\rm 1\,\mu G}\right)\,\left(\frac{E_{\rm Z,max}}{\rm 10^{20}\,{\rm eV}}\right)^{-1}\,.
\ee
}
The comparison between the timescales ($t_{\rm Z, acc}\lesssim t_{\rm Z, esc}\ll t_{lobe}$ for $Z\gtrsim 6$)  indicates that  nuclei can be accelerated up to $\sim 10^{20} {\rm eV} $ before they escape from the giant lobes.  It is worth noting that boron nuclei cannot be accelerated up to $10^{20}\,{\rm eV}$. The value of the gyromagnetic factor for $t_{\rm Z, acc} \simeq t_{\rm Z, esc}$ is $\eta\simeq  0.16 Z\, \left(\frac{\beta}{0.2}\right)   \left(\frac{B}{\rm 1\,\mu G}\right) \,\left(\frac{R}{\rm 100\, kpc}\right)\,\left(\frac{E_{\rm Z,max}}{\rm 10^{20}\,{\rm eV}}\right)^{-1}$, which lies in the Bohm diffusion limit for Z=6. This limit gives account that nuclei are efficiently accelerated in the lobes, thus being able to explain the UHECR flux in the direction around Cen A.\\ 
In order to determine the number and the chemical composition of UHECRs around Cen A, we estimate of the deflection angles through the Galactic and extragalactic magnetic fields.  In the Galactic disk,  the magnetic field  immersed in it is arranged  in a spiral with its lines frozen (from left to right) within the plane.  Hence, UHECRs can move randomly  to and fro in a vertical direction, filling the super Galactic plane \citep{2008PhyS...78d5901F}. The magnetic field in our Galaxy has been more studied that the extragalactic magnetic field. In our Galaxy, the strength of the field can be estimated, in principle, considering the pulsar rotation and the dispersion measurements  associated with the distribution of free electrons \citep{2003astro.ph..1598C, 2002astro.ph..7156C}. The orientation of the field which indicates that it increases towards the inner Galaxy has been inferred by simultaneous  studies of synchrotron radiation in the radio wavelengths, cosmic rays and $\gamma$-ray data.  Using the values of Galactic magnetic field in the halo and on the plane with a coherence length $\sim 1$ kpc, the average deflecting angles in our Galaxy  due to Galactic magnetic field in the halo is \citep{2008PhyS...78d5901F}
\be
\theta_{\rm Z,G}\gtrsim  1.3^\circ\, Z \, \left(\frac{E_{\rm A,th}}{58\, {\rm EeV}}\right)^{-1}\,  \left(\frac{B_G}{1\,\mu G}\right) \sqrt{\frac{L_G}{10\,{\rm kpc}}}\,\sqrt{\frac{l_{\rm c,G}}{{\rm kpc}}}\,, 
\ee
and on the plane is
\be
\theta_{\rm Z,G}\gtrsim  1.9^\circ\, Z \, \left(\frac{E_{\rm A,th}}{58\, {\rm EeV}}\right)^{-1}\,  \left(\frac{B_G}{4\, \mu G}\right) \sqrt{\frac{L_G}{20\,{\rm kpc}}}\,\sqrt{\frac{l_{\rm c,G}}{{\rm kpc}}}\,. 
\ee
It is worth emphasizing that Galactic magnetic field may create a strong shadowing effect on the true location of sources.   For an event traveling a distance of 3.8 Mpc, the mean-square deviation due to the extragalactic magnetic field ($B_{\rm EG}\sim$ 1 nG) with a coherence length of $l_{\rm c, EG}\sim$ 1 Mpc  is \citep{2017APh....89...14F} 
{\small
\be
\theta_{\rm Z, EG}\gtrsim  0.3^{\circ}Z\left(\frac{58\, {\rm EeV}}{E_{\rm A,th}}\right)^{-1} \left( \frac{B_{\rm EG}}{1\,{\rm nG}} \right)   \sqrt{\frac{L_{\rm EG}}{10\, {\rm Mpc}}}  \sqrt{\frac{l_{\rm c,EG}}{1\, {\rm Mpc}}}\,.
\ee
}
The average deflecting angles show that when the UHECRs go through a region endowed with a magnetic field,  heavy nuclei are deflected at larger angles than light nuclei. As expected, the average deflecting angles due to Galactic magnetic field on the plane is larger than that in the halo, and even larger than the extragalactic magnetic field.  Considering the extragalactic and the Galactic magnetic field on the plane,  the average deflecting angles are  $\theta_{\rm Z,G}\gtrsim2.1^\circ$, $4.4^\circ$, $13.2^\circ$, $15.2^\circ$ and $57.2^\circ$ for proton, helium, carbon,  nitrogen and iron nuclei, respectively. Therefore, UHE iron nuclei going through the Galactic plane can hardly arrive on Earth in a radius less than $\lesssim 15^\circ$, and UHE light nuclei such as  carbon/nitrogen nuclei can arrive.\\
On the other hand, using the values of magnetic fields  reported in Table 1, the magnetic luminosity becomes $L_B=\pi R^2 \Gamma^2\frac{B^2}{8\pi}\simeq 4.1\times 10^{43}\,\,{\rm erg\,\, s^{-1}\, \left(\frac{\beta}{0.2}\right)\,\left(\frac{R}{100\, {\rm kpc}}\right)^2\,  \left(\frac{B}{\rm \mu G}\right)^2\, \left(\frac{\Gamma}{1}\right)^2}$.  Considering the condition for accelerating UHE protons and nuclei, the magnetic luminosities are $L_B\gtrsim 2.0\times10^{44}\,{\rm erg\, s^{-1}}$, $5.0\times10^{43}\,{\rm erg\, s^{-1}}$, $5.5\times 10^{42}\,{\rm erg\, s^{-1}}$, $4.1\times 10^{42}\,{\rm erg\, s^{-1}}$  and $2.9\times10^{41}\, {\rm erg\, s^{-1}}$ for proton, helium, carbon, nitrogen and iron nuclei, respectively. Comparing the value of the magnetic luminosity found in Cen A, protons and He nuclei from this radio galaxy would be again excluded in the chemical composition of UHECRs.\\
Figure \ref{FMP} shows the energy-loss mean free path for photo-disintegration processes with CMB photons as a function of energy for helium, carbon, nitrogen and iron nuclei.  In addition, the mean free path of protons for photopion energy losses due to CMB photon field is displayed.   Based on the interactions with CMB photons, the most significant contribution of helium, carbon, nitrogen and iron nuclei/protons to UHECR composition  has as origin sources located to distances less than 5 Mpc,  70 Mpc and 150 - 400 Mpc, respectively.  It is important to highlight that any fraction of UHE helium nuclei detected on Earth must come from close sources ($\lesssim$ 5 Mpc), and  fractions of UHE iron nuclei and protons from sources longer than $\gtrsim$ 150 Mpc.  Considering UHECR events to distances less than $\lesssim$ 75 - 100 Mpc, a degree of anisotropy in a particular direction and a possible correlation with the chemical composition is expected.     Taking into consideration our analysis around the maximum nuclei/proton energy reached in the giant lobes with their timescales, the average deflecting angles, the magnetic-field luminosity  condition and the energy-loss mean free path for photo-disintegration processes, it  can be inferred that  inside a radius of  $\sim15^{\circ}$ centered around Cen A,   
UHE light nuclei such as carbon and nitrogen nuclei are the most promising candidates in the data collected by PAO. UHE iron nuclei and protons are excluded and UHE helium nuclei correspond to the highest flux at energies less than $\sim$ 40 EeV.   We show that the results obtained with our model are consistent with the results recently reported by  \cite{2014PhRvD..90l2006A} concerning to the UHECR composition at the highest energies.\\

\subsubsection{CR luminosities at lower energies}
We consider the 14 UHECR events reported by \cite{2015ApJ...804...15A} in a circular window radio of $15^\circ$ centered in Cen A in order to estimate the CR luminosities at distinct energies.    Column 4 in Table 2 shows the UHECR luminosity at 1 EeV for spectral indexes  $\alpha_A$= 2.2, 2.4 and 2.6, and  $N^{\rm obs}_{\rm cr}=14$.   In addition, columns 2 and 3 show the CR luminosities at 1 TeV and 100 PeV, respectively, which were calculated extrapolating the UHECR luminosity at these lower energies.    The number of UHECR events and the UHECR luminosity would have been increased by a factor of 1.5 - 2 if a deflection angle around 18$^\circ$ would have been considered.  It is worth noting that the UHECR luminosity  $\sim10^{40} - 10^{41}$ erg/s at 1 EeV corresponds to one of the principal sources of UHECR emissivity inside $\sim$ 10 Mpc \citep{PhysRevD.59.023002}. \\  
\begin{table}\renewcommand{\arraystretch}{1.45}\addtolength{\tabcolsep}{1.5pt}
\centering
\caption{Nuclei luminosity  at different energies.}\label{table2}
\begin{tabular}{ l c c c }
 \hline
 $\alpha_A$         &                &   $L_A\,({\rm erg/s})$        &   \\
                            & $1\, {\rm TeV}$  & $100\, {\rm PeV}$& $1\, {\rm EeV}$ \\

 \hline 
 \hline
\scriptsize{2.2}                              & \scriptsize{1.1$\times10^{41}$}  & \scriptsize{1.1$\times10^{40}$}    & \scriptsize{7.2$\times10^{39}$} \\
\scriptsize{2.4}                              & \scriptsize{2.3$\times10^{42}$}  &  \scriptsize{2.4$\times10^{40}$}    &  \scriptsize{9.4$\times10^{39}$} \\
\scriptsize{2.6}                              & \scriptsize{6.4$\times10^{43}$}  &  \scriptsize{6.4$\times10^{40}$}    &  \scriptsize{1.6$\times10^{40}$} \\
\hline
\end{tabular}
\end{table}
\begin{flushleft}
\scriptsize{
}
\end{flushleft}
It is important to highlight that the UHECR luminosities reported in Table 2 would have not changed significantly if the 2 events  collected from 2004 January  1$^{\rm th}$ until  2007 August 31 \citep{ 2007Sci...318..938P, 2008APh....29..188P}  would have been considered.  This relevant result shows that a total deflection angle of 15$^\circ$ around Cen A  can explain successfully the amount of  UHECRs coming from Cen A. It seems to be that Cen A has a constant rate of UHECR emission.\\
\subsection{High-energy $\gamma$-ray photons}
In agreement with our model, the $\gamma$-ray flux counterparts from the hadronic interactions can be estimated as follows.  The photopion efficiency for carbon nuclei  inside the giant lobes is $2.3\times 10^{-2}$ when the buoyancy arguments are considered, and  $1.2 \times 10^{-3}$ and $2.5 \times 10^{-3}$ when spectral arguments are taken into account for the southern and northern lobes, respectively.  Using the previous values of the photopion efficiencies and the CR luminosities at $\sim$1 TeV reported in Table \ref{table2},  the photon luminosities are calculated and reported in Table \ref{table3}.  Values reported in Table \ref{table3} without (with) round parenthesis are calculated considering the ages of the giant lobes estimated by  buoyancy (spectral) arguments.  Using the timescales for this hadronic process,  the CR luminosities at 1  TeV (see Table \ref{table2}) and the values reported in the literature (see Table \ref{table1}),  the total and CR energies as well as the total pressure in the giant lobes are reported in Table \ref{table3}. The values reported in this table for $\alpha_C\sim2.5$ and the ages of the lobes based on the spectral arguments are in accordance with those reported by \cite{2010Sci...328..725A, 2009MNRAS.393.1041H, 2014ApJ...783...44F}. Therefore, the $\gamma$-ray flux  coming from pion decay products contribute to the $\gamma$-ray flux reported by  \cite{2010Sci...328..725A} for $\alpha_C \sim 2.4$ and 2.5 - 2.6 when the buoyancy and spectral arguments are considered, respectively.\\

\begin{table*}
\centering
\caption{Derived quantities for the Lobes of Cen A}\label{table3}
\begin{tabular}{ l c c c c }
 \hline
 Parameters & \scriptsize{} & \small{Northern} &  &\small{Southern} \\
             & \scriptsize{$\alpha_A$} & \scriptsize{2.2\hspace{2.2cm}2.6} & &\scriptsize{2.2\hspace{2.2cm} 2.6} \\
 \hline 
 \hline
\scriptsize{$\gamma$-ray luminosity ($\times 10^{40}$ erg/s)}   & \scriptsize{L$_{\gamma}$}  &  \scriptsize{8.4$\times10^{-2}$ (9.2$\times10^{-3}$)\hspace{0.8cm}5.2 (48.4) } &  &\scriptsize{8.4$\times10^{-2}$(4.4$\times10^{-3}$) \hspace{0.6cm}0.6 (48.4)} \\
 \scriptsize{Pressure  ($\times 10^{-13}$ dyn cm$^{-2}$)}   & \scriptsize{P$_{nth}$}  &  \scriptsize{0.5 ($2.0\times 10^{-1}$)\hspace{0.9cm}$1.9\times 10^{2}$(18.3)} &  &\scriptsize{  1.1 ($1.8\times 10^{-1}$) \hspace{0.9cm}52.6 (9.1)} \\
\scriptsize{Total energy  ($\times 10^{58}$ erg)}               & \scriptsize{E$_{tot}$}  & \scriptsize{  0.6 ($2.4\times 10^{-1}$)\hspace{0.9cm}$2.4\times 10^{2}$ (22.5)} & & \scriptsize{  1.4 ($2.3\times 10^{-1}$) \hspace{0.9cm}64.8 (11.1)}   \\
\scriptsize{CR energy  ($\times 10^{58}$ erg)} & \scriptsize{E$_A$}  & \scriptsize{  0.5 ($3.9\times 10^{-2}$)\hspace{0.9cm}$2.4\times 10^{2}$ (22.3)} &  &\scriptsize{ 1.3 ($1.9\times 10^{-2}$) \hspace{0.9cm}64.7 (10.9)}\\
\hline
\end{tabular}
\end{table*}

\begin{figure}
\vspace{0.4cm}
{\centering
\resizebox{0.5\textwidth}{0.3\textheight}
{\includegraphics{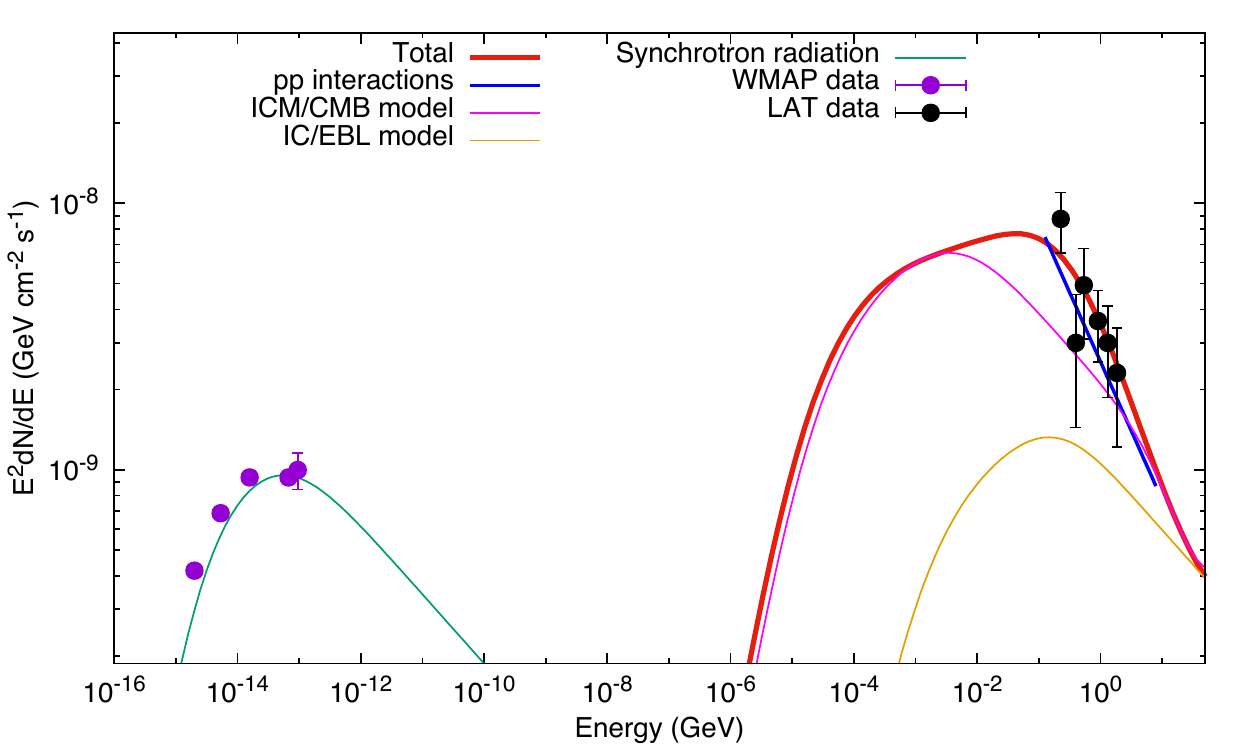}}
\resizebox{0.5\textwidth}{0.3\textheight}
{\includegraphics{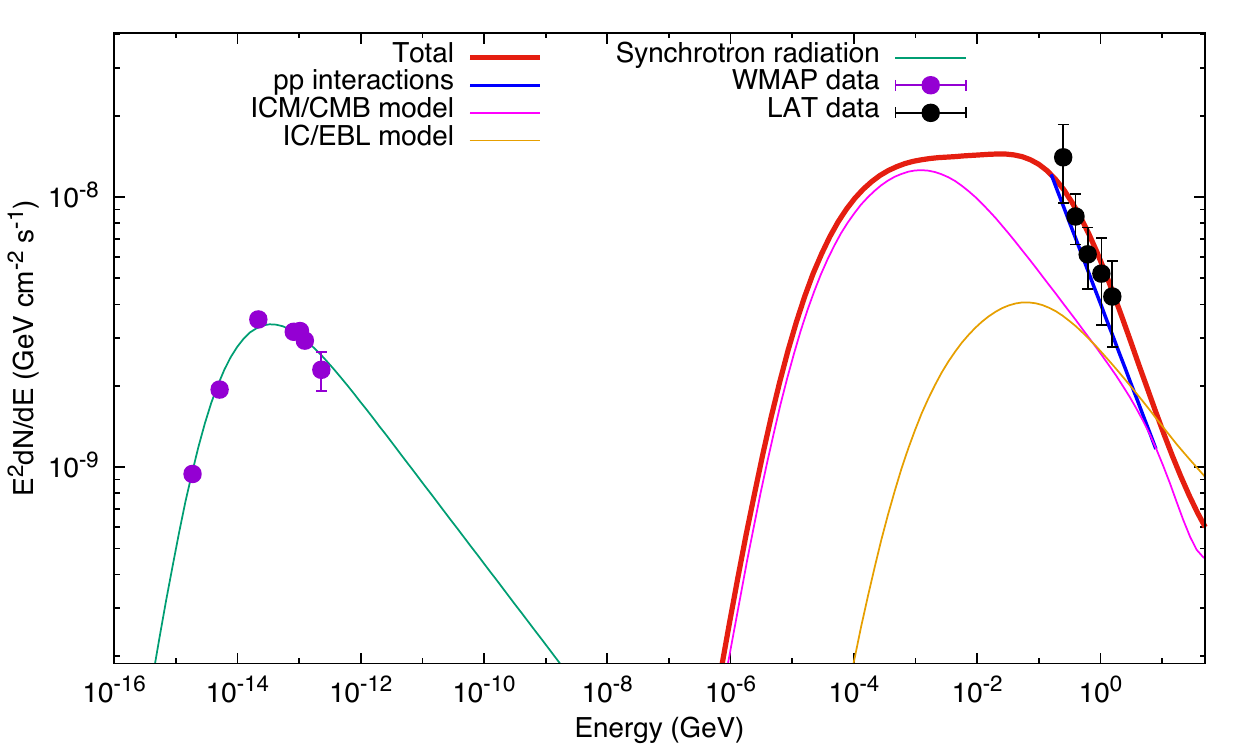}}
}
\caption{Fits of the broadband SED of the northern (top) and southern (bottom) extended lobes. We use  synchrotron emission  to describe WMAP data \citep{2009ApJS..180..225H,2003ApJS..148...39P, 2009MNRAS.393.1041H, 2010Sci...328..725A}  and IC/CMB, IC/EBL and hadronic interactions to model LAT data \citep{2010Sci...328..725A}.}
\label{gamma_Lobes}
\end{figure}

In order to describe the spectral energy distribution (SED) of the northern and southern lobes, the lepto-hadronic model showed in \citet{2014ApJ...783...44F} is used with accelerating carbon nuclei instead of protons.    Figure \ref{gamma_Lobes} shows the fit of the SED of the northern (left panel) and southern (right panel) lobes.   The Wilkinson Microwave Anisotropy Probe  data  \citep[WMAP;][]{2009ApJS..180..225H,2003ApJS..148...39P} at radio wavelengths (22 to 94 GHz) are described with synchrotron emission.   The Fermi-LAT data  \citep{2010Sci...328..725A}  are interpreted as the superposition of inverse Compton with CMB and EBL photon fields and hadronic interactions. The typical values of CMB and EBL photon densities and energies were used from \cite{2013avhe.book..225D}, and the values of electron Lorentz factors from \citet{2014ApJ...783...44F}. The contribution of pion decay products from hadronic interactions were calculated  considering the photopion efficiency based on spectral arguments and  the CR luminosities normalized with the number of UHECRs, and extrapolated to energies as low as 1 TeV.  In addition, we consider the photon indexes reported by Fermi-LAT instrument; $2.52^{+0.16}_{-0.19}$  for the northern lobe and $2.60^{+0.14}_{-0.15}$  for  the southern lobe which correspond to the CR luminosities for carbon nuclei of $L_A=1.37^{+0.09}_{-0.10}\times 10^{43}\, {\rm erg\, s^{-1}}$  for the northern and  $0.62^{+0.04}_{-0.04}\times 10^{43}\, {\rm erg\, s^{-1}}$   for southern lobe.  Figure  \ref{gamma_Lobes} displays that the neutral pion decay products coming from hadronic interactions between the relativistic carbon nuclei and thermal particles inside the giant lobes  are consistent  with GeV $\gamma$-ray flux.
% 
%
%
%

%{\centering
%\resizebox*{0.5\textwidth}{0.3\textheight}
%{\includegraphics{A_eff_v1.pdf}}
%}

\subsection{PeV neutrinos}
\begin{figure}
\vspace{0.4cm}
{\centering
\resizebox*{0.5\textwidth}{0.3\textheight}
{\includegraphics{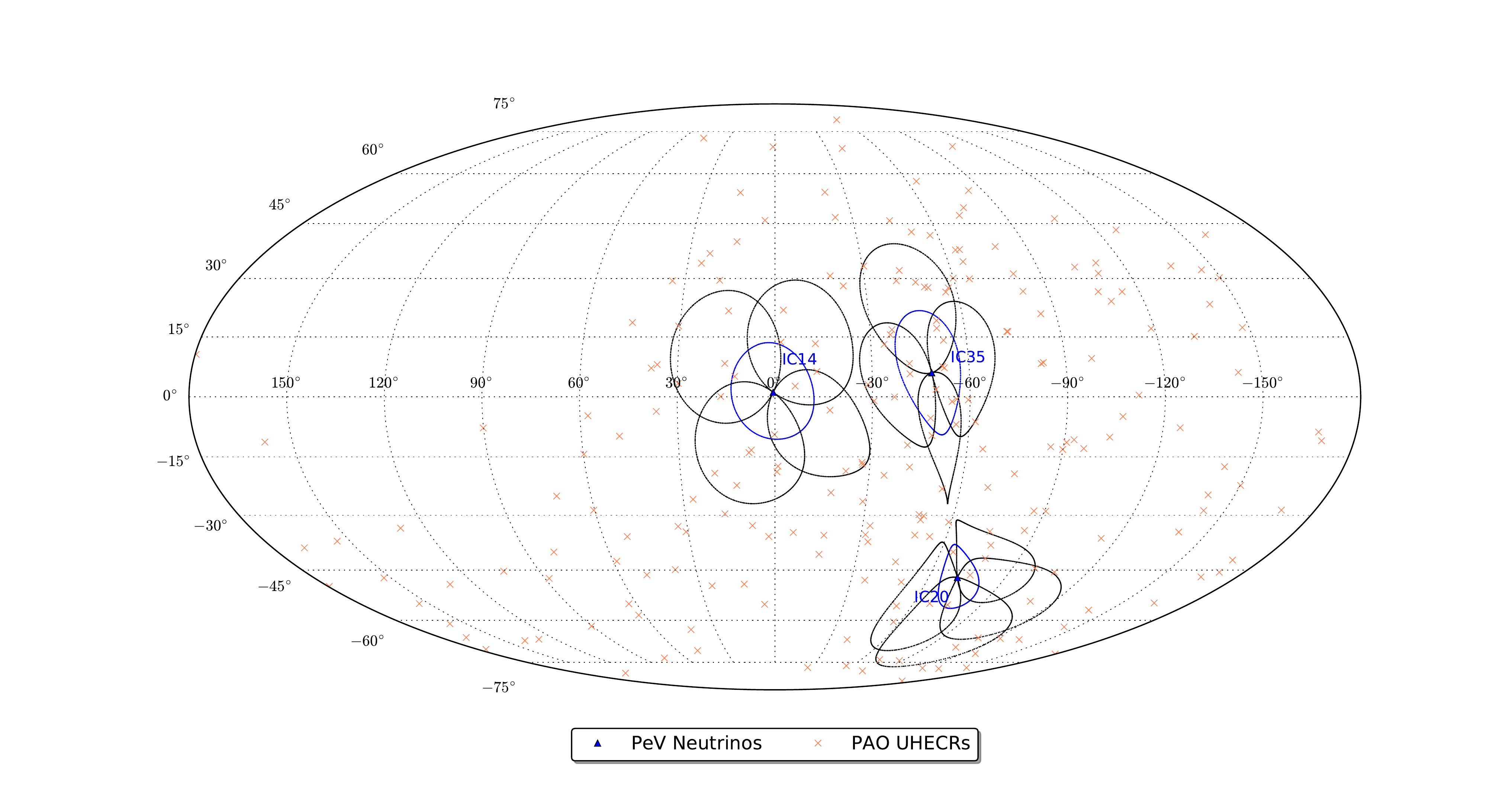}}
}
\caption{Sky-map with the PeV neutrino events  detected by IceCube (HESE catalog; blue triangles) and the UHECR events collected by PAO \citep[red crosses; ][]{2015ApJ...804...15A}.  The blue contour represents the median angular errors of the neutrino events.    In order to correlate the PeV neutrinos  with the number of UHECRs around them  black circles of 15$^\circ$ radii are shown.}   
\label{PeVNevents}
\end{figure}

\begin{figure}
\vspace{0.4cm}
{\centering
\resizebox*{0.4\textwidth}{0.2\textheight}
{\includegraphics{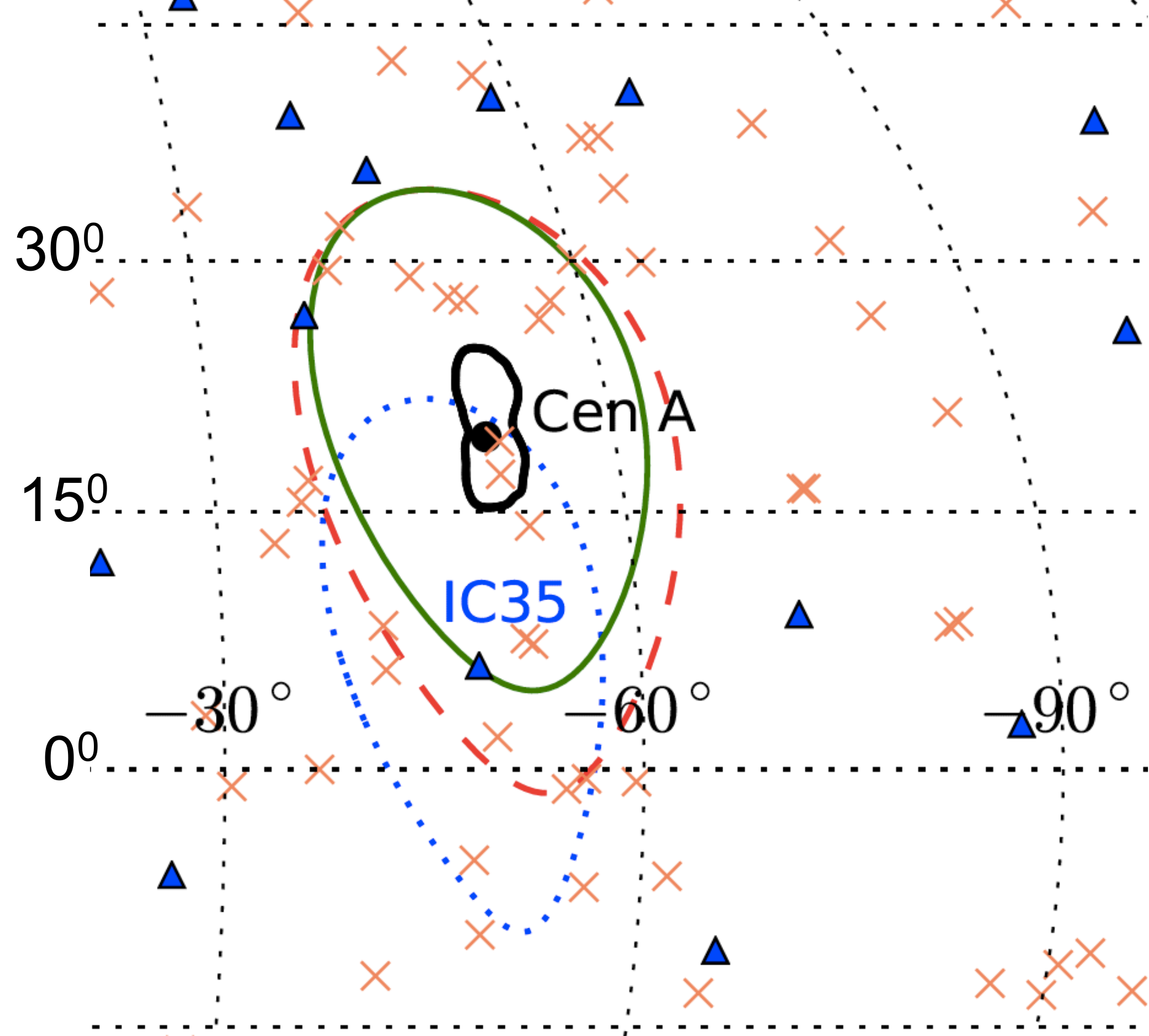}}
}% 
\caption{A blow-up of the sky-map around Cen A (black point) with a green and a red circle of 15$^\circ$ around Cen A and its southern lobe, respectively, are exhibited.  The UHECR and neutrino events are shown in red crosses and blue triangles, respectively. The blue dotted line is the median angular error of IC35 neutrino event. The contour of the large-scale structure of the giant lobes  at 1.4 GHz was considered \citep[black line;][]{2009MNRAS.393.1041H}.}
\label{SkymapZoom}
\end{figure}

Sky-map with the PeV neutrino events from the HESE catalog (IC14, IC20 and IC35) and the UHECR events collected by PAO \citep[red crosses; ][]{2015ApJ...804...15A} is shown in Figure \ref{PeVNevents}.  The blue contour represents the median angular errors of the neutrino events.    In order to correlate the PeV neutrinos  with the number of UHECRs around them  black circles of 15$^\circ$ radii are shown. In this figure is shown  that  IC14 event is associated with 2  events in a circle centered at $\sim$ (DEC:-15$^\circ$, R.A.:-6$^\circ$), IC20 event is associated with 2  events in a circle centered at $\sim$ (DEC:--90$^\circ$, R.A.:52$^\circ$) and  IC35  is associated with 10  events in a circle centered at  $\sim$ (DEC:-41$^\circ$, R.A.:15$^\circ$).\\ 
A blow-up of the sky-map around Cen A (black point) with a green and a red circle of 15$^\circ$ around Cen A and its southern lobe, respectively, is shown in Figure \ref{SkymapZoom}.  The UHECR and neutrino events are shown in red crosses and blue triangles, respectively. The blue dotted line is the median angular error of IC35 neutrino event. The contour of the large-scale structure of the giant lobes  at 1.4 GHz was considered \citep[black line;][]{2009MNRAS.393.1041H}. This figure shows that IC35 event is inside the circular region of 15$^\circ$ centered in the southern giant lobe.  Taking into consideration the green circle can be observed that the IC35 event would be outside  of the circular region of 15$^\circ$ centered in the northern giant lobe.   Therefore,  the southern giant lobe will only be analyzed around the IC35 event. In this case, the number of UHECRs associated to both the southern giant lobe and the IC35 event is 10 and then, the CR luminosity linked to these events are 0.7 times those reported in Table 2.\\ 
In order to analyzed the IC35 event, we consider the carbon nuclei which is the most favorable candidate to explain the chemical composition and also the  accelerated and diffusion timescales for $E_C=300$ PeV.  The 300-PeV carbon nuclei are accelerated in a  timescale of $t_{\rm C, acc}\simeq  2.1\times 10^{-4}\,{\rm Myr}$,  and then they diffuse through the giant lobes, and later escape in a timescale of  $t_{\rm C, esc}\simeq 7.2\times 10^2\,{\rm Myr}$.   By comparing the escape timescale of the  300-PeV carbon nuclei with the ages of the giant lobes (using the buoyancy and spectral arguments),  it can be noticed that carbon nuclei cannot escape from the giant lobes if their ages are obtained by spectral arguments. Otherwise, they could escape only when the ages are those estimated by the buoyancy arguments.  Hereafter, the ages of the giant lobes estimated by the buoyancy arguments will be considered.   We consider the hadronic interactions inside and outside the lobes. 
The hadronic interactions in the inner and giant lobes will be analyzed. \\

\subsubsection{Giant lobes}
The carbon nuclei with energies less than $\sim$ 100 EeV as shown in eqs. (\ref{tac}) and (\ref{tdif}) interact within the giant lobes with an average density proton of $n_p=10^{-4}\,{\rm cm^{-3}}$ (see Table \ref{table4}).   In this case,  the efficiency of photopion is $1.1\times 10^{-2}$, and therefore, the number of events is much less than one, as shown through the two-dotted-dashed magenta line  in Figure \ref{Nev}. This line in each panel displays the number of events as a function of the spectral index of carbon nuclei for $\alpha_\nu$ = 2.2, 2.3, 2.4 and 2.5.  The maximum number of $\sim$ 2-PeV neutrinos expected from the southern giant lobe when the carbon nuclei interact with a thermal particle density  is $\sim$ 0.08 for $\alpha_{\rm C}$ = 2.8.\\
\\
The 300-PeV carbon nuclei in the direction of the IC35 event  interact with CMB and EBL photons.    For the photo-hadronic interactions with EBL photons, we take into account the  infrared (IR), optical and ultraviolet (UV) photon densities which are  $6.3\times 10^{-1}\,{\rm cm^{-3}}$,  $2.1\times 10^{-2}\,{\rm cm^{-3}}$ and $3.2\times 10^{-4}\,{\rm cm^{-3}}$ for $\epsilon_{\rm \gamma,br}\simeq$ 10$^{-2}$, 1 and 4 eV (see eq. \ref{pgamma}), respectively. Due to ultra-relativistic carbon nuclei interacting with CMB photons ($\epsilon_{\rm \gamma,br}\simeq 2.34\times 10^{-4}\,{\rm eV}$) produce a neutrino flux which peaks at $\sim$ EeV \citep[e.g.][]{2016ApJ...830...81F}, these interactions will not be considered.    Dashed blue line in Figure \ref{Nev}  shows the number of events as a function of the spectral index of carbon nuclei when IR photons are considered for $\alpha_\nu$ = 2.2, 2.3, 2.4 and 2.5. In this case, the maximum number of events expected is $8\times 10^{-3}$ for $\alpha_{\rm C}$ = 2.8. The numbers of neutrino events from the p$\gamma$ interactions with UV and optical photons are not shown in this figure because these are much less than $10^{-5}$.

\begin{figure}
\vspace{0.4cm}
{\centering
\resizebox*{0.5\textwidth}{0.5\textheight}
{\includegraphics{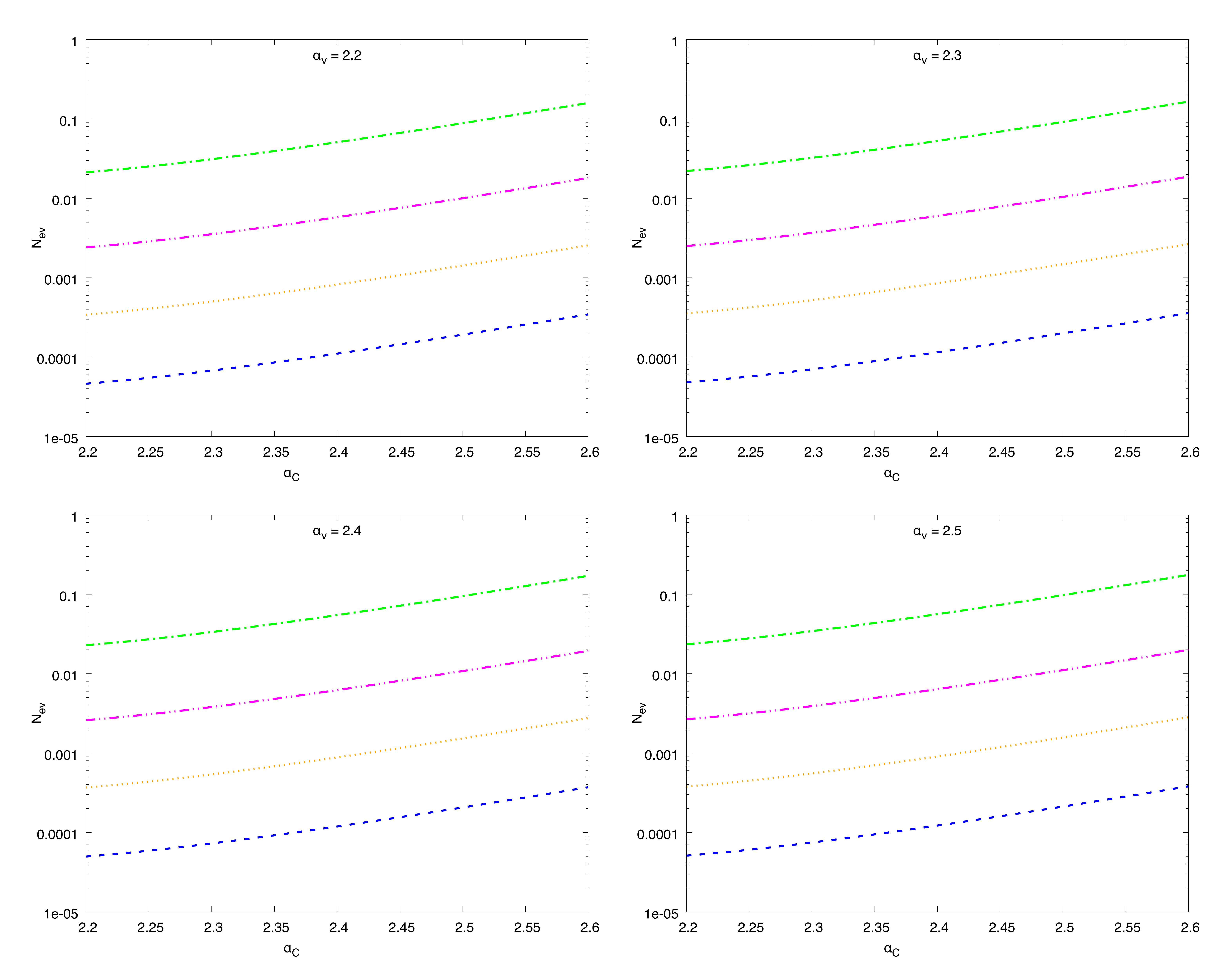}}
}
\caption{Number of 2-PeV neutrino events as a function of  carbon-nuclei and neutrino  spectral indexes.  This figure shows the neutrino contribution produced by ultra-relativistic carbon nuclei interacting with Galaxies  in the path to Earth (dot-dashed green line),  with thermal density inside the giant lobes (double dot-dashed magenta line),  with synchrotron photons close to the core (dotted orange line)  and with IR photons (dashed blue line).}   
\label{Nev}
\end{figure} 

\subsubsection{Inner lobes}
Analysis of  the  Chandra/ACIS-I and XMM-Newton observations of X-ray emission from the inner lobes and ambient medium showed an X-ray filament consisted of five spatially resolved X-ray knots in the northern radio lobe \citep {2009ApJ...698.2036K} and a bright  X-ray enhancement along the edge of southwest radio lobe \citep{2003ApJ...592..129K, 2007ApJ...665.1129K}.   The origin of the X-ray knots were explained as the result of cold gas heated by a direct interaction with the jet.  The X-ray emission along the edge of the southwest radio lobe was modelled as a thin and hot shell X-ray emitting plasma. Due to the temperature and density of the gas in the hot shell is much higher than the ambient medium, then the inflation of the southwest radio lobe is driving a strong shock into ambient medium. Table \ref{table4} shows the densities and lifetime of the knots, the hot shell and the ambient medium in the northern and southwest inner lobes, respectively.\\
\begin{table}
\centering
\caption{Parameters of the knots, the hot shell and the ambient medium  in the northern and southwest inner lobes, respectively.}\label{table4}
\begin{tabular}{ l c c c c }
 \hline
Region & \scriptsize{} & \scriptsize{Density} &\scriptsize{Lifetime} &\scriptsize{Photopion Efficiency}\\
	   & \scriptsize{} & \scriptsize{10$^{-2}$(cm$^{-3}$)} &\scriptsize{(Myr)} &\scriptsize{} \\
 \hline 
 \hline
\scriptsize{$^{b}$N1}   & \scriptsize{}  &  \scriptsize{0.8} & \scriptsize{3.7} & \scriptsize{$7.2\times10^{-3}$}\\
\scriptsize{$^{b}$N2}   & \scriptsize{}  &  \scriptsize{1.1} & \scriptsize{2.6} & \scriptsize{$6.8\times10^{-3}$}\\
\scriptsize{$^{b}$N3}   & \scriptsize{}  &  \scriptsize{1.1} & \scriptsize{2.4} & \scriptsize{$6.2\times10^{-3}$}\\
\scriptsize{$^{b}$N4}   & \scriptsize{}  &  \scriptsize{1.2} & \scriptsize{3.1} & \scriptsize{$8.6\times10^{-3}$}\\
\scriptsize{$^{b}$N5}   & \scriptsize{}  &  \scriptsize{2.7} & \scriptsize{0.6} & \scriptsize{$3.5\times10^{-3}$}\\
\scriptsize{$^{c}$A. medium}   & \scriptsize{}  &  \scriptsize{0.17} & \scriptsize{$\sim(10^2 - 10^3$)$^a$} & \scriptsize{$3.9\times10^{-2}$}\\
\scriptsize{$^{c}$Shell}   & \scriptsize{}  &  \scriptsize{2.0 - 2.2} & \scriptsize{$2.0$\,} & \scriptsize{$9.2\times10^{-3}$}\\
\hline
\end{tabular}
\begin{flushleft}
\scriptsize{
Notes.\\ $^a$ This value is reported considering the cooling time for the hot ambient medium in the central regions of elliptical galaxies.\\
$^b$ Region located in the northern inner Lobe.\\
$^c$Region located in the southwest inner Lobe.\\
\textbf{References}.  \cite{2009ApJ...698.2036K}; \cite{2003ApJ...592..129K,2007ApJ...665.1129K}; \cite{2009MNRAS.395.1999C}}
\end{flushleft}
\end{table}
\\
Taking into account the radii of inner lobes $\sim$ 10 kpc  \citep{2003ApJ...592..129K, 2007ApJ...665.1129K},  the maximum energy that carbon nuclei can reach by stochastic processes inside the inner lobes  is $\sim$ 10 EeV. Therefore, the acceleration and the diffusion timescales for 10-EeV carbon nuclei are $t_{\rm C, acc}\sim 0.16$ Myr and $t_{\rm C, esc}\sim 0.24$ Myr.   For this case, the luminosities cannot be normalized with the UHECRs found around the IC35 event due to the ample discrepancy between the maximum energy that ultra-relativistic carbon nuclei reached in the inner lobes and the minimum UHECR detected by PAO.  Because of that both timescales are similar $t_{\rm C, acc}\sim t_{\rm C, esc}$, 10-EeV carbon nuclei can be accelerated and then, escape from the inner lobes.   In order to analyze if the accelerated carbon nuclei have the potential to produce a 2-PeV neutrino, the accelerated timescale of $t_{\rm C, acc}\simeq  2.1\times 10^{-4}\,{\rm Myr}$ and the diffusion timescale of  $t_{\rm C, esc}\simeq 7.2\,\times 10^2\,\,{\rm Myr}$ for the 300-PeV carbon nuclei were calculated. Our analysis is divided for the northern and  southwest inner lobes.\\
\begin{itemize}
\item Northern inner lobe. The acceleration and diffusion timescales of the carbon nuclei and the lifetimes of knots,  $t_{\rm C, acc}\ll  t_{\rm lt,k} \ll t_{\rm C, esc}$,  indicate that these nuclei can be accelerated  but cannot escape from the knots.  Here, $ t_{\rm lt,k}$ is used to define the lifetime of the knots.  Taking into account the thermal density, the photopion production efficiency lies in the range of $2.4\times 10^{-3}\leq {\rm f_{\rm C\,p}} \leq 8.6\times 10^{-3}$  (see Table \ref{table3}).  In this case, unrealistic luminosities larger than $10^{44}<$ erg/s at $300$ PeV are necessary to produce a 2-PeV neutrino event.
\item  Southwest inner lobe.    Similarly, the timescales $t_{\rm C, acc}\ll t_{\rm lt,s}\ll t_{\rm C,esc}$,  show that these nuclei can be accelerated up $\sim$ 300 PeV but cannot escape from the hot shell and the ambient medium.  Here, $ t_{\rm lt,s}$ is used to define the lifetime of the hot shell and the ambient medium.  Considering the thermal density of the hot shell and the ambient medium,  the photopion production efficiency lies in the range $9.1\times 10^{-3}\leq {\rm f_{\rm C\,p}} \leq 3.9\times 10^{-2}$ (see Table \ref{table3}).    Similarly, for this inner lobe unrealistic luminosities again larger than $10^{44}<$ erg/s at $300$ PeV are necessary to produce a 2-PeV neutrino event.
\end{itemize}
\section{Analysis outside the extended Lobes}
\begin{figure}
\vspace{0.4cm}
{\centering
\resizebox*{0.4\textwidth}{0.3\textheight}
{\includegraphics{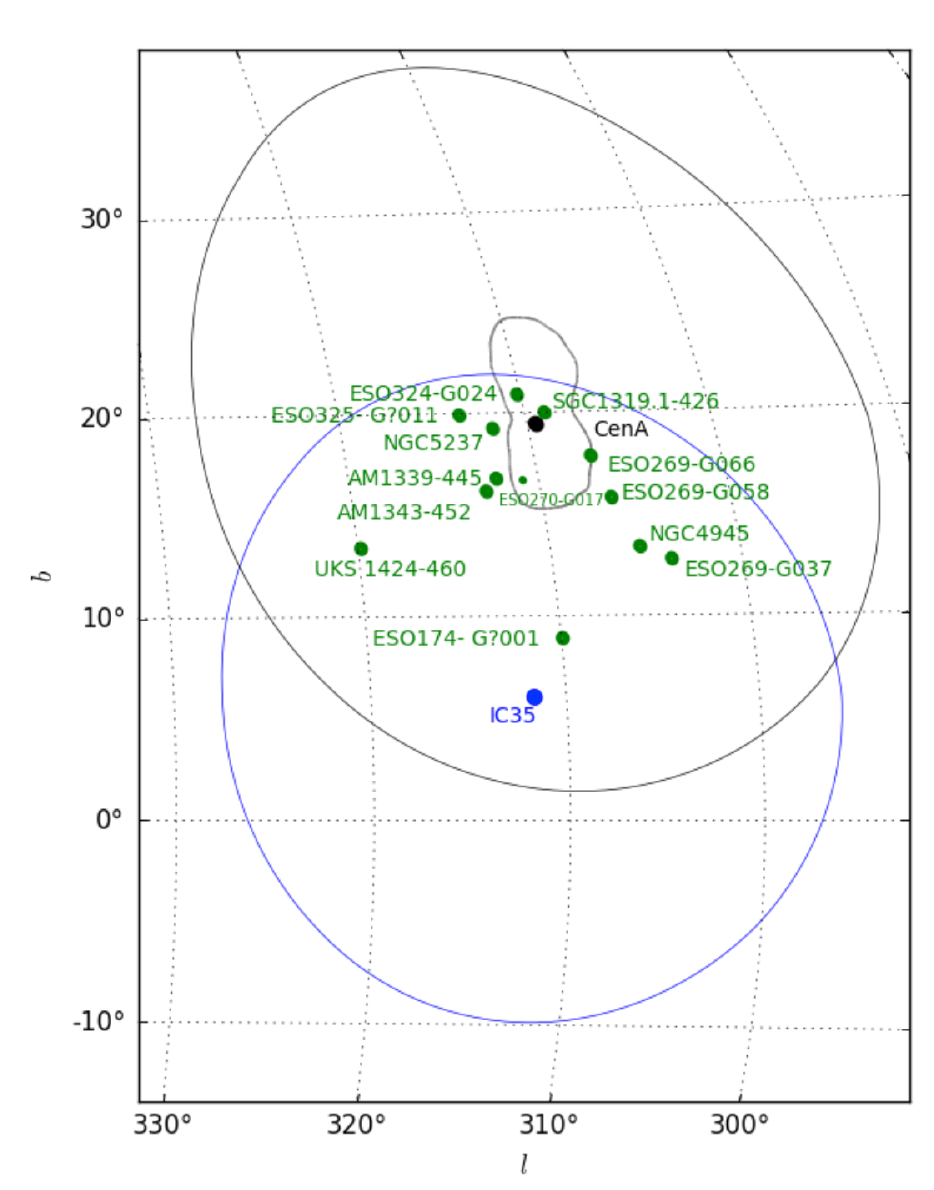}}
}
\caption{The closest Galaxies located at distances less than $\sim$ 4 Mpc and around the lobes of Cen A and the IC35 event.}   
\label{galaxies}
\end{figure}

Considering the  diffusion timescale and the ages of the giant lobes,  carbon nuclei with energies less than 300 PeV can hardly escape from the giant lobes and interact out of them. Therefore, neutrinos with energies less than 2 PeV cannot be expected outside the giant lobes.  On the other hand, carbon nuclei with energies larger than $\sim$ 300 PeV can escape from the giant lobes and interact around them.   Once these ultra-relativistic carbon nuclei abandon the giant lobes and are directed towards the Earth, they can interact in their path to Earth (in our galaxy or out of it).  The optical depths for these hadronic interactions in their paths to Earth are
\be
\tau\simeq 8.4\times 10^{-2}\,  \left(\frac{\rm n_p}{\rm cm^{-3}} \right)\,     \left(\frac{l}{100\,{\rm kpc}}\right)\,.
\ee
Here, $l$ is the average distance  where the interactions occur.  Considering the average distance of $\lesssim 4$ Mpc between Cen A and Earth  with density  $\sim 10^{-3}\,{\rm cm^{-3}}$, the optical depth becomes  $3.5\times 10^{-3}$. Similarly, considering the average radius of our Galaxy of 20 kpc  with density of $\sim 1\, {\rm cm^{-3}}$ \citep{2009ARA&A..47...27K},  the optical depth is $\sim 10^{-2}$.  Therefore, the probability so that relativistic carbon nuclei interact outside or/and inside our galaxy is very low.\\
\begin{table}
\centering\renewcommand{\arraystretch}{1.2}\addtolength{\tabcolsep}{-2pt}
\caption{Densities and magnetic field of Galaxies around  IC35 event and southern lobe}\label{table5}
\begin{tabular}{ l c c c c c c}
 \hline
Galaxy & \scriptsize{Type} & \scriptsize{Density} &\scriptsize{B}& \scriptsize{$t_{Cp}$} & \scriptsize{$t_{\rm diff}$} & \scriptsize{}\\
	   & \scriptsize{} & \scriptsize{(cm$^{-3}$)} &\scriptsize{($\mu$G)}& \scriptsize{(Myear)}&\scriptsize{(Myear)}& \scriptsize{}\\
 \hline 
 \hline
\scriptsize{ESO324-G024}   & \scriptsize{dIrr}  &  \scriptsize{15} &\scriptsize{39.9}&\scriptsize{3.2}& \scriptsize{4.5}&  \scriptsize{(1)} \\
\scriptsize{ESO325-G011}   & \scriptsize{dE}  &  \scriptsize{1.39} &\scriptsize{7.6}&\scriptsize{34.5}& \scriptsize{0.3}& \scriptsize{(3)} \\
\scriptsize{NGC5237}   & \scriptsize{dIrr}  &  \scriptsize{2.14} &\scriptsize{10.2}&\scriptsize{22.5}& \scriptsize{0.6}& \scriptsize{(3)} \\
\scriptsize{AM1339-445}   & \scriptsize{dE}  &  \scriptsize{$<10^{-2}$} &\scriptsize{0.2}&\scriptsize{$4.7 \times 10^3$}& \scriptsize{$1.5 \times 10^{-2}$}& \scriptsize{(1)} \\
\scriptsize{AM1343-452}   & \scriptsize{dE}  &  \scriptsize{$<10^{-2}$} &\scriptsize{0.2}&\scriptsize{$4.7 \times 10^3$}& \scriptsize{$1.5 \times 10^{-2}$}& \scriptsize{(1)} \\
\scriptsize{UKS1424-460}   & \scriptsize{dIrr}  &  \scriptsize{0.18} &\scriptsize{1.8}&\scriptsize{$2.7 \times 10^{2}$}& \scriptsize{$1.0 \times 10^{-1}$}& \scriptsize{(3)} \\
\scriptsize{ESO174-G001}   & \scriptsize{dE}  &  \scriptsize{$<10^{-2}$} &\scriptsize{0.2}&\scriptsize{$4.7 \times 10^3$}& \scriptsize{$1.5 \times 10^{-2}$}& \scriptsize{(1)}\\
\scriptsize{SGC1319.1-426}   & \scriptsize{dE}  &  \scriptsize{$<10^{-2}$} &\scriptsize{0.2}&\scriptsize{$4.7 \times 10^3$}& \scriptsize{$1.5 \times 10^{-2}$}& \scriptsize{(1)} \\
\scriptsize{ESO269-G066}   & \scriptsize{dIrr}  &  \scriptsize{$<10^{-2}$} &\scriptsize{0.2}&\scriptsize{$4.7 \times 10^3$}& \scriptsize{$1.5 \times 10^{-2}$}& \scriptsize{(1)} \\
\scriptsize{ESO269-G058}   & \scriptsize{dIrr}  &  \scriptsize{0.26} &\scriptsize{2.3}&\scriptsize{$1.8 \times 10^2$}& \scriptsize{$1.3 \times 10^{-1}$}& \scriptsize{(3)} \\
\scriptsize{NGC4945}   & \scriptsize{SB(s)cd}  &  \scriptsize{140} &\scriptsize{$1.9 \times 10^{2}$}&\scriptsize{0.4}& \scriptsize{7.2}& \scriptsize{(2)} \\
\scriptsize{ESO269-G037}   & \scriptsize{dIrr}  &  \scriptsize{$<10^{-2}$} &\scriptsize{0.2}&\scriptsize{$4.7 \times 10^3$}& \scriptsize{$1.5 \times 10^{-2}$}& \scriptsize{(1)} \\
\hline
\end{tabular}
\begin{flushleft}
\scriptsize{
\textbf{References}.  $^{(1)}$ \cite{2007AJ....133..261B}; $^{(2)}$\cite{2007ApJ...670..116C}; $^{(3)}$\cite{2009AJ....138.1037C} }
\end{flushleft}
\end{table}
Figure \ref{galaxies} shows the galaxies located at distances less than $\lesssim$ 4 Mpc and around the giant lobes of Cen A and the IC35 event.  Assuming that carbon nuclei could arrive inside these galaxies,  the photopion efficiencies are calculated through the hadronic cooling and diffuse timescales.  In this case, the efficiency can be written as 
\be\label{fpp2}
f_{\rm C\, p}=1-\exp(-t_{\rm C,esc}/t_{\rm C\,p})\,,
\ee
where  $t^{-1}_{C\,p}\simeq  \sigma_{C\,p}\,n_p$.    Using the relation between the gas contained in a Galaxy and the star-formation rate, which is given by \citep{1998ApJ...498..541K} 	
\be
\Sigma_{SFR}= A_{SFR} \left(\frac{\Sigma_{\rm gas}}{1\, {\rm M_\odot\, pc^{-2}}}\right)^{\alpha_k} {\rm M_\odot\, yr^{-1}\, kpc^{-2}}\,,
\ee
with $[A_{SFR},\,\alpha_k]= [(2.5\pm0.7) \times10^{-4},\, 1.4\pm0.15$],  we can estimate the gas density in the case of Dwarf Irregular (dIrr) galaxies inside a radius of $\sim$ 1 kpc for a height of $\sim$  500 pc.  In this case, the density can be written as
\be
n_g\simeq 24 \left(\frac{\rm SFR}{M_\odot \, {\rm yr^{-1}}}\right)^{0.71}\, \left( \frac{\rm R}{{\rm kpc}} \right)^{-1.43}\, \left(\frac{l}{500\, {\rm  pc}} \right)^{-1} {\rm cm^{-3}}\,.
\ee
It is worth noting that in the case of Dwarf Elliptic (dE) galaxies,  a same upper limit of gas density can be used \citep{2007AJ....133..261B}.\\

Table  \ref{table5} shows the galaxies located at distances less than $\sim$ 4 Mpc and around the lobes and the IC35 event. This table exhibits the name of galaxies (column 1), type (column 2),  the density (column 3), magnetic field (column 4), interaction timescale (column 5) and diffusion time (column 6) associated to them.  These galaxies in general are dE-type that are gas-poor and dIrr-type that are gas-rich.   The strengths of the magnetic field were calculated using the relation \citep{2006ApJ...645..186T}
\begin{equation}
B \simeq 5.864 \; \mu G \; \left(\frac{n_g}{\rm cm^{-3}} \right)^{0.7} \left(\frac{l}{500 \, \rm pc} \right)^{0.7}\,,
\end{equation}
and the diffuse timescales were calculated in accordance to eq. (\ref{tdif})  for carbon nuclei with energies of $\sim$ 300 PeV.  From eq. (\ref{fpp2}) and the values of $t_{\rm C\,p}$ and $t_{\rm C,diff}$ reported in Table \ref{table5}, it is shown that the  photopion efficiency around unity could occur when the relativistic carbon nuclei interact with the material around ESO324-G024 and NGC4945.  Although a rigorous work must be done about these gaxies, it is out of the scope the paper.   In this scenario, these galaxies  together with a minimum luminosity of $\sim$ 10$^{40}$-10$^{41}\, {\rm erg\,s^{-1}}$ as obtained in this work are the most promising sources to create a PeV neutrino. Considering the average density of the galaxies around Cen A, we compute the number of PeV neutrinos (dot-dashed green line in Figure \ref{Nev}) produced by ultra-relativistic carbon nuclei interacting with these galaxies in the path to Earth.     A similar work  for our Galaxy with a density of 1 cm$^{-3}$  was done.  It indicates that if the 300-PeV carbon nuclei survive the deflection due to intergalactic magnetic field and reach our Galaxy,  the photopion efficiency becomes $\sim2\times 10^{-2}$ which is very low to produce a PeV neutrino.\\
On the other hand, it is important to highlight again that as the IC35 event is outside of the circular window of 15$^\circ$ around northern giant lobe, the  analysis through this paper  was done  for the southern giant lobe. However,  the two neutrino events, IC48  and IC49, with energies $104.7^{+13.5}_{-10.2}$ TeV and $59.9^{+8.3}_{-7.9}$, respectively, could be spatially correlated with the northern giant  lobe.   In this model, these neutrino events would not be produced by relativistic carbon nuclei with energies of $\sim$ 6 and 3 PeV, respectively, because these nuclei would escape from the northern giant lobes in timescales of tens of Gyr, which are larger than the ages of this lobe.\\ 
\section{Conclusions}
The analysis done in this paper has been based on the PeV-neutrino event (IC35) reported in the HESE catalog,  the UHECR events collected by PAO between 2004 January 1$^{\rm th}$ and 2014 March 31 and the morphology of the lobes of Cen A.
\\
\\
Considering  the UHECR timescales in the inner lobes,  the maximum energies that protons and the heaviest nuclei can reach are  0.4 and  23.8 EeV, respectively, which are below the energy window observed by PAO.  Using the timescales in the giant lobes of Cen A, the UHE proton, helium and lithium cannot be accelerated in this case up to energies larger than $\geq$ 58 EeV. Therefore, giant lobes in Cen A are the only places around this source where nuclei heavier than beryllium can be accelerated at UHEs.\\
\\
Taking into consideration the extragalactic and the Galactic magnetic fields,  the average deflecting angles are larger than  $\theta_{\rm Z,G}\gtrsim2.1^\circ$, $4.4^\circ$, $13.2^\circ$, $15.2^\circ$ and $57.2^\circ$ for proton, helium, carbon,  nitrogen and iron nuclei, respectively. Therefore, UHE nuclei heavier than oxygen accelerated in Cen A can hardly arrive on Earth in a radius less than $\lesssim 15^\circ$. \\
\\
We showed that the magnetic luminosities for proton, helium, carbon, nitrogen and iron nuclei are $L_B\gtrsim 2.0\times10^{44}\,{\rm erg\, s^{-1}}$, $5.0\times10^{43}\,{\rm erg\, s^{-1}}$, $5.5\times 10^{42}\,{\rm erg\, s^{-1}}$, $4.1\times 10^{42}\,{\rm erg\, s^{-1}}$  and $2.9\times10^{41}\, {\rm erg\, s^{-1}}$, respectively. Comparing the magnetic luminosity found after modelling the broadband SED in giant lobes of Cen A, protons and He nuclei from this radio galaxy would be again excluded as UHECRs accelerated by this radio galaxy.\\
\\
Based on the photo-disintegration processes that nuclei undergo in their paths,  we showed that the most significant contribution of helium, carbon and iron nuclei to the UHECR composition  has as origin sources located to distances less than 5 Mpc,  70 Mpc and 150 - 400 Mpc, respectively.  It shows that any fraction of UHE helium nuclei detected on Earth must come from sources $\lesssim$ 5 Mpc, and  fractions of UHE iron nuclei from sources further than $\gtrsim$ 150 Mpc.  Considering UHECR events to distances less than $\lesssim$ 75 - 100 Mpc, a degree of anisotropy in a particular direction and a possible correlation with the chemical composition is expected.  For protons, we consider the photopion energy losses and found that a large fraction would come from sources further than $\gtrsim$ 150 Mpc. \\ 
\\
Based on  i) the maximum energy that nuclei can reached in the giant lobes, ii) the average deflecting angles that nuclei can undergo due to Galactic and extragalactic magnetic fields, iii)  the magnetic-field luminosity condition of UHECRs in the giant lobes and iv) the energy-loss mean free path for photo-disintegration processes,  we showed  that  inside a radius of  $\sim15^{\circ}$ centered around the radio galaxy Cen A,    UHE light nuclei such as carbon and nitrogen are the most promising candidates,  and UHE iron nuclei and protons are excluded.  This is in agreement with  the results recently reported by  \cite{2014PhRvD..90l2006A} about the UHECR composition at the highest energies.\\
\\
Considering the 14 events reported by PAO inside a circular window of radius 15$^\circ$ centered on Cen A, we normalize the UHECR luminosities using  a simple  power  law.   With the values of parameters reported in the literature and found from modelling the broadband SED, we have showed some acceleration mechanisms that could accelerate nuclei  inside the giant lobes up to energies as high as $10^{20}$ eV.   Taking into consideration the CR luminosity extrapolated at 1 TeV we have described the faint $\gamma$-ray fluxes reported by the Fermi Collaboration assuming relativistic carbon nuclei.\\ 
\\
Focusing on the circular window of radius 15$^\circ$ centered in the southern giant lobe and  the median angular error of the IC35 neutrino event, 10 UHECRs were spatially correlated.  By normalizing the luminosity  at 300 PeV with these ten events,  the number of neutrinos was computing through photohadronic and hadronic interactions inside and outside the giant lobes, assuming carbon nuclei.  The IceCube effective area at Cen A declination was used.  We showed that although hadronic interactions are a more efficient processes than photo-hadronic interactions, none of them can generate a 2-PeV neutrino inside the inner and/or giant lobes.  Otherwise,
the probability to produce a 2 PeV neutrino through the interaction of carbon nuclei in their path to Earth results very low.  Therefore, we do not find enough evidence to associate the IC35 event with the UHECR events reported by PAO. Finally,  it is important to highlight that the increasing statistics by the IceCube neutrino telescope and PAO  will provide more evidence of our scenario.\\  
\section*{Acknowledgements}
We  thank Charles Dermer and Maria Petropoulou for useful discussions.  This work was supported by UNAM PAPIIT grant IA102917.
\end{document}